\documentclass{vgtc}                          

\graphicspath{{figures/}{pictures/}{images/}{./}} 

\usepackage{times}                     

\newcommand{\sysname}{SonoCraftAR}

\makeatletter
\newcommand\footnoteref[1]{\protected@xdef\@thefnmark{\ref{#1}}\@footnotemark}
\makeatother

\usepackage{tabu}                      
\usepackage{booktabs}                  
\usepackage{lipsum}                    
\usepackage{mwe}                       
\usepackage{soul} 

\usepackage{mathptmx}                  

\usepackage{listings}

\lstdefinestyle{mystyle}{
    basicstyle=\sffamily\footnotesize,
    breakatwhitespace=false,         
    breaklines=true,                 
    keepspaces=true,                 
    showspaces=false,                
    showstringspaces=false,
    showtabs=false,                  
    tabsize=2
}
\onlineid{0}

\vgtccategory{Research}

\vgtcinsertpkg

\title{\sysname{}: Towards Supporting Personalized Authoring of Sound-Reactive AR Interfaces by Deaf and Hard of Hearing Users}

\author{Jaewook Lee\\ %
     \scriptsize University of Washington %
\and Davin Win Kyi\\ %
     \scriptsize University of Washington %
\and Leejun Kim\\ %
     \scriptsize University of Washington %
\and Jenny Peng\\ %
     \scriptsize University of Washington %
\and Gagyeom Lim\\ %
     \scriptsize Dong-A University %
\and Jeremy Zhengqi Huang\\ %
     \scriptsize University of Michigan %
\and Dhruv Jain\\ %
     \scriptsize University of Michigan %
\and Jon E. Froehlich\\ %
     \scriptsize University of Washington %
}

\teaser{
  \centering
  \includegraphics[width=\textwidth]{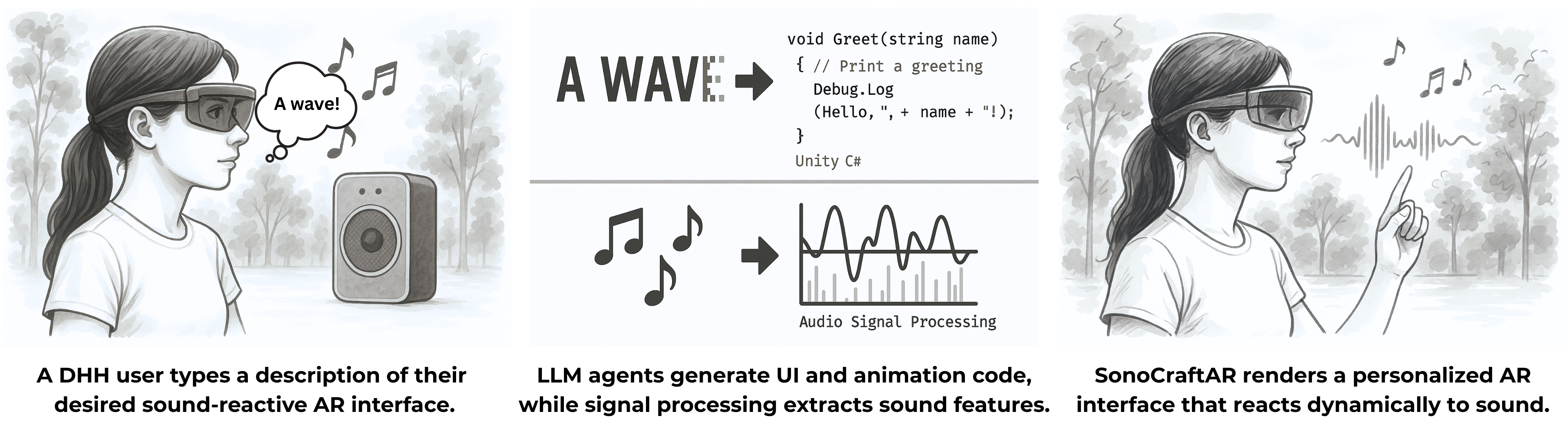}
  \caption{\textit{\sysname{}} empowers Deaf and hard‑of‑hearing (DHH) users to author personalized, sound‑reactive interfaces in wearable augmented reality (AR). Users type a desired UI description, which a multi‑agent LLM pipeline converts into procedurally generated, animated visualizations. Our prototype system performs real-time audio signal processing to extract features (\textit{i.e.,} dominant frequency), which drive the animations and make the visualizations respond dynamically to sound.}
  \label{fig:teaser}
}

\abstract{
Augmented reality (AR) has shown promise for supporting Deaf and hard‑of‑hearing (DHH) individuals by captioning speech and visualizing environmental sounds, yet existing systems do not allow users to create personalized sound visualizations. We present \textit{\sysname{}}, a proof‑of‑concept prototype that empowers DHH users to author custom sound‑reactive AR interfaces using typed natural language input. \sysname{} integrates real‑time audio signal processing with a multi‑agent LLM pipeline that procedurally generates animated 2D interfaces via a vector graphics library. The system extracts the dominant frequency of incoming audio and maps it to visual properties such as size and color, making the visualizations respond dynamically to sound. This early exploration demonstrates the feasibility of open‑ended sound‑reactive AR interface authoring and discusses future opportunities for personalized, AI‑assisted tools to improve sound accessibility.
}


\begin{document}



\maketitle

\section{Introduction}
Augmented reality (AR) has shown promise as an assistive technology for people who are Deaf and hard of hearing (DHH), including for real-time captioning~\cite{guo2020holosound, huang2025soundweaver, jain2018exploring, jain2018mobileconversation, miller2017dhhstudents, olwal2020wearablesubtitles, peng2018speechbubbles} and visualizing non-speech sounds~\cite{guo2020holosound, huang2025soundweaver, jain2015headmounted}. While research has explored how to personalize sound recognition models using advances in deep learning~\cite{bragg2016personalizedapp, goodman2021towardspersonalization, goodman2025spectra, jain2022protosound}, limited prior work exists on personalizing assistive sound visualizations. One recent system, \textit{SoundWeaver}, automatically adjusts AR interfaces based on a user’s surroundings and intent~\cite{huang2025soundweaver}; however, it does not let users actively author new visualizations (\textit{e.g.,} by typing ``\textit{a wave}'') and instead focuses on automatic UI adaptation. While these systems offer valuable support, they remain largely preconfigured and do not yet allow DHH users to create their own personalized, sound-reactive AR interfaces.

Empowering disabled individuals to design their own tools is a crucial step toward inclusive technology~\cite{herskovitz2023diy, potluri2018codetalk}. Recent advances in generative AI open new possibilities for supporting user-driven AR creation. For example, several systems enable XR scene composition through natural language using large language models (LLMs)~\cite{torre2024llmr, giunchi2024dreamcodevr, lee2025imaginatear, vachha2025dreamcrafter, zhang2024vrcopilot}. Some also support simple movements~\cite{huang2023animation, leiva2021rapido, leiva2020pronto, xia2023realitycanvas}, though generating animations---especially those that respond to real-world sensory input---remains challenging. Critically, no prior XR authoring tool has explored open-ended AR interface creation for making sound more accessible.

\begin{figure*}[hbt!]
  \centering
  \includegraphics[width=\linewidth]
  {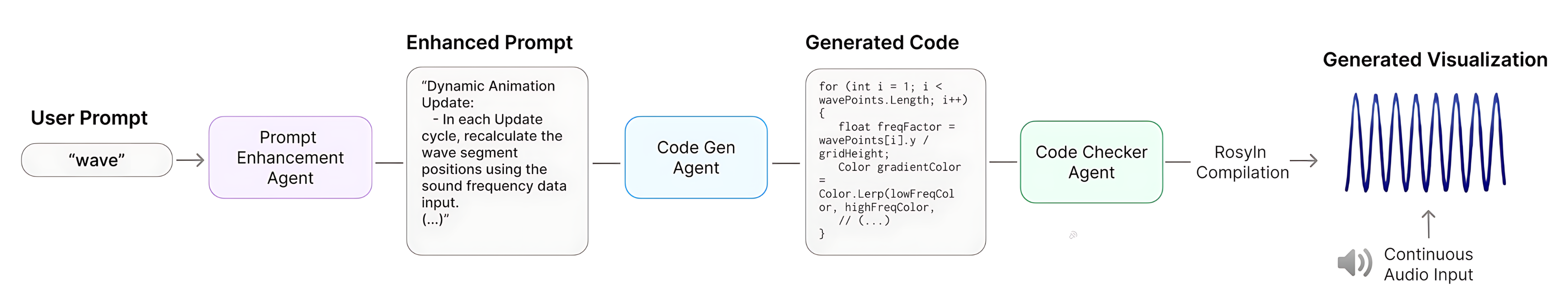}
  \caption{System overview of \sysname{}. A typed user prompt is first expanded by the \textit{Prompt Enhancement} agent into structured implementation guidelines. The \textit{Code Generation} agent then produces a Unity C\# script that uses the Shapes vector graphics library, which is checked for compilation errors by the \textit{Code Checker} agent. The finalized code script is compiled at runtime with Roslyn and rendered in AR. A real‑time audio processing server continuously computes the dominant frequency of incoming sound, which drives the visualization’s animations.}
  \label{fig:system-diagram}
\end{figure*}

We introduce \textit{\sysname{}}, a novel proof-of-concept prototype that empowers DHH users to author personalized, sound-reactive AR interfaces through typed natural language input. Built for the \textit{Microsoft HoloLens 2}\footnote{\label{HoloLens}\url{https://learn.microsoft.com/en-us/hololens/hololens2-hardware}}, \sysname{} combines real-time audio signal processing with LLM agents to generate responsive 2D AR UIs that visualize ambient sound. To support flexible authoring, our system includes three \textit{o3}~\cite{GPTo3} agents: a \textit{Prompt Enhancement} agent that supplements user input with UI accessibility and design guidelines; a \textit{Code Generation} agent that outputs a \textit{Unity}\footnote{\label{Unity}\url{https://unity.com}} C\# script using code documentation and examples; and a \textit{Code Checker} agent that detects and resolves compile-time errors. UIs are rendered using \textit{Roslyn}\footnote{\label{Roslyn}\url{https://assetstore.unity.com/packages/tools/integration/roslyn-c-runtime-compiler-142753}} for runtime code compilation and built with the \textit{Shapes}\footnote{\label{Shapes}\url{https://assetstore.unity.com/packages/tools/particles-effects/shapes-173167}} vector graphics library, which provides fine-grained control over visual properties for animations. To equip the LLM with relevant prior knowledge, we crawled the Shapes documentation and included it as part of the LLM system prompt. Interfaces respond to sound features such as dominant frequency by adjusting properties like shape size, color, and saturation. Together, these components enable open-ended authoring of sound-reactive AR interfaces tailored to DHH users’ preferences and needs.

In this workshop paper, we describe the \sysname{} system and discuss its limitations and future opportunities for enabling hyper-personalized, sound-reactive UI authoring for DHH users.

\begin{figure*}[hbt!]
  \centering
  \includegraphics[width=\linewidth]
  {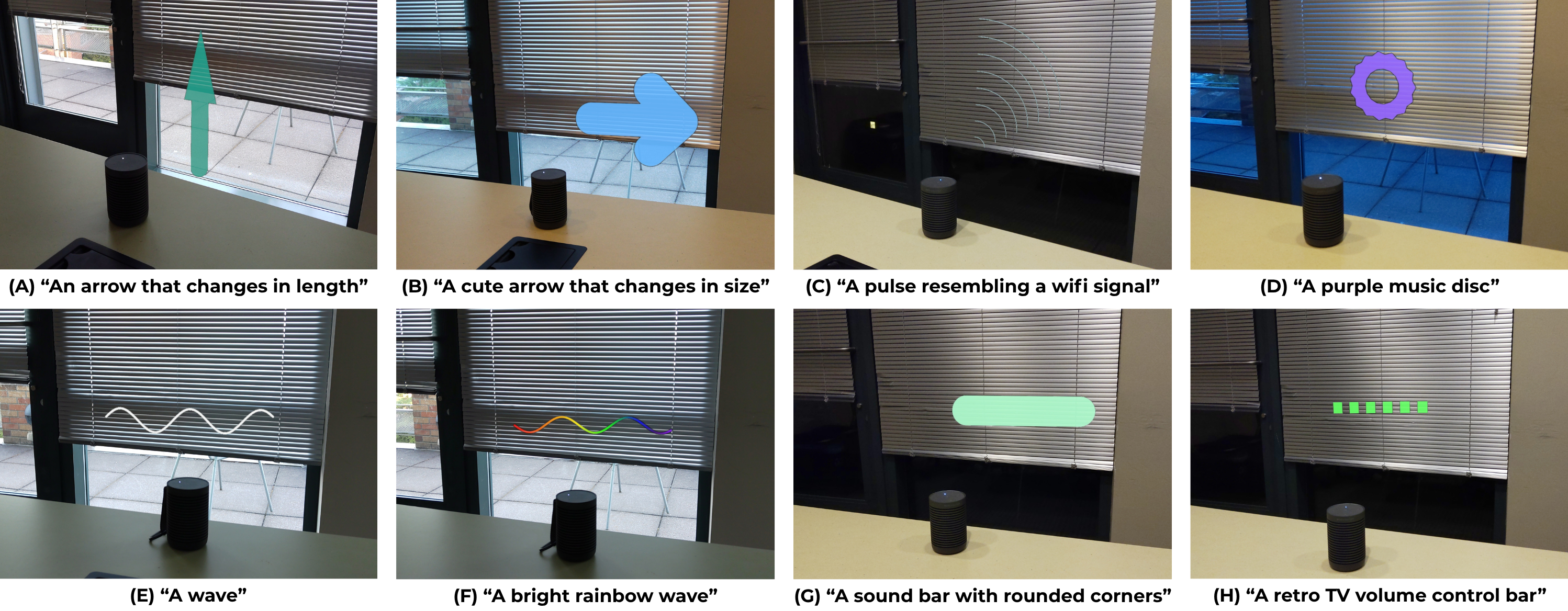}
  \caption{Eight example sound‑reactive interfaces created with \sysname{}. The designs include arrows, waves, pulsing arcs, sound bars, and more. Examples A–C were inspired by GlassEar~\cite{jain2015headmounted}, while D–H showcase ideas proposed by the research team. These visualizations react to the dominant frequency of music playing through a speaker.}
  \label{fig:design-space}
\end{figure*}

\section{The \sysname{} System}
We begin by describing how we enabled LLMs to utilize a vector graphics library, followed by how LLMs are prompted to generate code that defines the structure and behavior of AR sound interfaces, which are then made sound-reactive through audio signal processing. By enabling Deaf and hard of hearing (DHH) users to author their own sound visualizations, \textit{\sysname{}} facilitates more personalized representations---allowing users to perceive and interpret sound in ways that reflect their individual preferences.

\subsection{Enabling LLM-Based Procedural UI Rendering}
To support open-ended, sound-reactive AR interface generation, we required a rendering system that not only produces visually compelling UIs but also allows fine-grained control over individual components for animation. While recent work has explored generating 2D assets using diffusion-based models~\cite{podell2023sdxlimprovinglatentdiffusion, rombach2022highresolutionimagesynthesislatent}, such outputs are poorly suited for responding to real-time sensory input like sound. In contrast, vector graphics enable procedural rendering, allowing properties such as color, size, and position to be modified dynamically at runtime based on audio features.

We selected \textit{Shapes}\footnoteref{Shapes}, a real-time vector graphics library for \textit{Unity}\footnoteref{Unity}, to support code-driven UI rendering. Shapes enables procedural drawing of primitives such as lines, polygons, and arcs, which can be composed into more complex 2D elements. This allowed us to treat each UI as a collection of smaller, independently controllable components that can be updated in real time based on audio input---producing animated visualizations. For example, our system can pulse element size in sync with dominant frequency.

However, writing correct Shapes-based code requires familiarity with its custom C\# API, which differs from Unity’s default rendering primitives. To make this interface usable by LLMs, we crawled the official Shapes documentation\footnote{\url{https://acegikmo.com/shapes/docs}} using \textit{BeautifulSoup}\footnote{\url{https://www.crummy.com/software/BeautifulSoup/}} to extract its HTML source, then converted it to markdown using \textit{html-to-markdown}\footnote{\url{https://html-to-markdown.com}}. The resulting document was reviewed by two researchers and included in the LLM system prompt as a reference guide. This approach enabled LLMs to generate code that accurately uses the Shapes API.

\subsection{System Implementation}
\sysname{} is implemented on the \textit{Microsoft HoloLens 2}\footnoteref{HoloLens} using Unity 2022.3.52f1 and the \textit{Mixed Reality Toolkit (MRTK)} 2.8.3\footnote{\url{https://learn.microsoft.com/en-us/windows/mixed-reality/mrtk-unity/mrtk2}}. The system employs a multi-agent LLM pipeline to expand user prompts, generate sound-reactive UI code, and check for compile-time errors. All prompts used are included in the Appendix. The resulting interface is rendered at runtime and responds dynamically to the dominant frequency of incoming sound via continuous audio signal processing. Because the HoloLens does not support runtime script compilation, \sysname{} includes a Windows laptop running the Unity Editor, which compiles and executes the generated scripts. We then use Holographic Remoting\footnote{\url{https://learn.microsoft.com/en-us/windows/mixed-reality/develop/native/holographic-remoting-overview}} to stream the resulting holographic content to the HoloLens 2 in real time for viewing. We detail system components below.

\textbf{Understanding User Input.} Modern speech transcription tools are typically trained on speech data from non-DHH speakers and often struggle to accurately transcribe speech from DHH users~\cite{Glasser2017}. To address this limitation, \sysname{} supports typed natural language input, allowing users to provide prompts with reduced risk of recognition errors.

\textbf{Expanding User Prompt.} Users provide typed descriptions of desired sound visualizations. This input is passed to the \textit{Prompt Enhancement} agent, which uses OpenAI's \textit{o3}~\cite{GPTo3} to convert it into structured implementation instructions. The enhanced prompt adds details on how the visualization should look and animate (\textit{e.g.,} pulsing size, shifting color, varying saturation or thickness) and specifies how to assemble relevant Unity and Shapes API methods. This enriched prompt lets the next agent focus on generating accurate code rather than simultaneously making design decisions---a separation that empirically reduced hallucinations.

\textbf{Script Generation.} The enhanced prompt is passed to the \textit{Code Generation} agent, which uses one‑shot prompting with the Shapes library documentation, a short Unity C\# checklist (\textit{e.g.,} namespaces), a template, and an example output to produce an appropriate script. Before doing so, it determines whether the user’s query is for a new output or an iteration. The agent receives the previous prompt and code but is instructed to use this context only for follow‑up queries, where it edits the existing visualization; otherwise, it ignores this context and creates a new script. The generated code also includes a function callable by the \textit{Roslyn}\footnoteref{Roslyn} runtime compiler that takes real‑time sound data as input, allowing the UI to dynamically respond to audio features.

\textbf{Checking for Errors.} After a script is generated, it is compiled, and any resulting errors are passed to the \textit{Code Checker} agent along with the code and user prompt. Leveraging the Shapes documentation and a list of common Unity C\# errors (\textit{e.g.,} floats must end with `f'), the agent fixes compile‑time issues and verifies that the code matches the user’s request. This final step improves reliability by catching errors missed during generation and producing a corrected script ready for deployment.

\textbf{Rendering in AR.} Finally, the generated script is compiled at runtime with Roslyn and rendered in AR through Holographic Remoting. Roslyn continuously supplies real‑time sound data (\textit{i.e.,} the dominant frequency value) to the script, enabling the visualization to react dynamically to incoming audio.

\textbf{Audio Signal Processing.} To extract sound features in real time, \sysname{} runs a Python server that samples stereo audio on a Windows laptop at 48 kHz and processes it in 100 ms non-overlapping chunks. The signal is first converted to mono and passed through a Hanning window to reduce spectral leakage. A real-valued Fast Fourier Transform (FFT) implemented with NumPy\footnote{\url{https://numpy.org/doc/stable/reference/generated/numpy.fft.rfft.html}} is then applied to compute frequency-domain magnitudes, from which the dominant frequency---the peak‑magnitude component within 20-8000 Hz (covering most speech and environmental sounds while excluding low-frequency rumble and high-frequency noise)---is extracted. This value is normalized to a 0–10 scale using logarithmic mapping, enabling smoother visual scaling in UI animations. The result is sent to the Unity Editor via a WebSocket connection. To keep the proof-of-concept system simple while demonstrating sound-reactivity, we chose to represent only the dominant frequency, which is quick to compute and often matches the perceived pitch of tonal sounds like music.

\textbf{Latency.} To evaluate runtime performance, we ran the system 50 times using the prompt ``\textit{a sound wave}'' and observed an average generation time of $36.6\pm13.9$ seconds to produce a sound-reactive visualization.

\textbf{Example Interaction.} Suppose a user types ``\textit{a wave}''. \sysname{} could generate a white waveform visualization (Figure~\ref{fig:design-space}E) that reacts to music playing from a speaker. They can then follow up with ``\textit{make it rainbow‑colored}'', to which our system could update the wave with a rainbow gradient (Figure~\ref{fig:design-space}F).

\subsection{Example Creations} 
We demonstrate several sound‑reactive AR interfaces authored with \sysname{}, including an arrow that changes in length or width, pulsing arcs, a rainbow‑colored wave, and a retro‑style TV volume bar. Some of these recreate UIs proposed in \textit{GlassEar}~\cite{jain2015headmounted}, which outlined a design space for AR sound visualizations. Each example reacts dynamically to Beyoncé’s \textit{Love on Top}, chosen for its varying dominant frequency, playing from a nearby speaker. Figure~\ref{fig:design-space} shows all example creations.

\section{Discussion and Future Work}
\sysname{} is an early proof‑of‑concept that explores how DHH users can author personalized, sound‑reactive AR interfaces. In this section, we outline the system’s limitations and highlight opportunities for future work.

\subsection{Personalization vs. Effective Sound Representation}
There is a potential trade‑off across open‑ended personalization, visual fidelity, and the effectiveness of sound representation. A curated set of professionally designed visualizations could ensure high visual quality, convey sound features more clearly by following established design principles, and reduce user burden through ready‑made options. In contrast, \sysname{} allows DHH users to create entirely new visualizations on the fly, offering greater flexibility and creative freedom. However, this open‑ended approach can lead to inconsistent quality or unclear sound‑to‑visual mappings, as generative models are inherently less predictable. This unpredictability, though, also creates opportunities for unexpected and novel designs that curated approaches may miss.

Future work could examine this tension through user studies to better understand when and what DHH users author with an open-ended system, as well as how DHH users balance effectiveness and creative control. A hybrid approach may be promising---for example, providing curated templates as starting points while still enabling free‑form creation~\cite{amershi2019haiguideline}. Additionally, AI could play a more active role in co‑creation by suggesting ideas when users are unsure, or by asking clarifying questions (\textit{e.g.,} ``\textit{Do you want to add color or patterns to the wave?}'')~\cite{torre2024llmr, lee2025imaginatear}. Users could also be given a ``\textit{creativity slider}'' to control how adventurous they want the AI to be~\cite{lee2025imaginatear}. Such a system could empower users to design more personalized visualizations without sacrificing usability or visual quality.

\subsection{Limitations and Future Work}
Beyond the need to study the balance between personalization and effectiveness and the potential for human‑AI co‑creation, several additional directions remain open for future work:

\textbf{Supporting Multiple Sound Features and Sources.} Our current system visualizes only the dominant frequency of a single audio input. Future work could incorporate additional features (\textit{e.g.,} loudness, pitch contours, timbre) and support multiple sound sources through sound localization or speaker diarization. This would enable users to author multiple sound‑reactive UIs at once, attach them to specific objects, and reposition them as desired.

\textbf{Expanding Visualization and Interaction Modalities.} Future iterations could integrate other visualization libraries beyond Shapes, such as p5.js\footnote{\url{https://p5js.org}} or d3.js\footnote{\url{https://d3js.org}}, to broaden creative possibilities. For example, p5.xr\footnote{\url{https://github.com/stalgiag/p5.xr}} is an add-on that enables running p5 sketches in AR via WebXR\footnote{\url{https://immersiveweb.dev}}. We could also support richer interaction modalities---such as a voice interface for DHH users comfortable with speaking, or sketch-based input for drawing visualizations or marking elements for modification---to make authoring more flexible and accessible.

\textbf{Leveraging Generative Model Advances.} While we use LLMs to generate code, vision-language models (VLMs) could be incorporated to better understand good and bad visualizations from image or video examples, helping the system follow established design principles. Increasing the context window could also allow users to iterate across multiple rounds of edits while maintaining richer conversational history.

\textbf{More robust error recovery.} Although we include a Code Checker agent, occasional errors persist. In our current implementation, simple follow‑up prompts (\textit{e.g.,} ``\textit{I don’t see anything}'') can trigger self‑correction, but more sophisticated recovery methods are possible. For example, recording a short video of the generated UI and analyzing it with a VLM could help detect visual errors---such as an arrow’s triangle tip being positioned too far from or overlapping the shaft. Undo functionality could further give users control when AI outputs are undesirable.

\textbf{User Study.} Future iterations of \sysname{} should be evaluated through user studies with DHH participants to better understand the opportunities and challenges of authoring personalized sound‑reactive interfaces. A key question could be how DHH users will interact with \sysname{}, particularly how they might refine visualizations until they are just right.

\textbf{Lower latency.} Finally, reducing latency will be critical for usability of an in-situ authoring system.

\section{Conclusion}
In this workshop paper, we introduced \sysname{}, a proof‑of‑concept prototype that empowers DHH users to author personalized, sound‑reactive AR interfaces. With advances in generative AI, it is now possible to design systems that allow users to create assistive tools on the fly. We hope this work inspires further exploration of AI‑ and AR‑based systems that empower people with disabilities to design tools for their own needs and preferences---particularly custom sound visualizations.


\bibliographystyle{abbrv-doi}

\bibliography{paper}
\end{document}


\onecolumn
\begingroup
\raggedright
{\LARGE\bfseries Supplementary Materials for \sysname{}\par}
\vspace{1.5\baselineskip}
\endgroup
\noindent This supplement lists all prompts used by the LLM agents in \sysname{}.
\appendix
\section{Shapes Documentation}
\label{lst:shapes-documentation}
Due to its length (875 lines), the full Shapes documentation is provided \href{https://drive.google.com/file/d/1Wu-W2PDjSVL7SVyiJ0E5AUpWiCITFx31/view?usp=sharing}{\underline{here}}. Steps to reproduce it are described in the main text.
\section{Prompt Enhancement Agent Prompt}
\begin{lstlisting}[caption={Prompt used in \textit{Prompt Enhancement} to specify visualization appearance and animation behavior using Unity and Shapes API.},label={lst:prompt-enhancement},frame=single]
**Role**
   *** You are a prompt engineer for a Unity system that helps generate UI elements at runtime given a user prompt. This UI is designed to help Deaf and hard of hearing users visually perceive environmental sounds. Your task is to expand the user prompt into a clear, numbered list of implementation steps for a MonoBehaviour C# script. This list will be passed to another AI agent responsible for generating the actual code. Please generate a prompt that enhances the user prompt with the following requirements below.

**Enhancement Requirements**
    *** The enhanced prompt will include instructions to create visualizations that change dynamically with one of the sound data attributes that has been provided: volume, pitch, frequency, direction, and distance.
    *** Make sure that the UI is not too large or too small.
    *** Make sure to describe how the overall UI dynamically moves and which parts of the UI are moving
        **** For example, if the user prompt is: "a sound bar", the enhanced prompt could say something like "the fill amount of the sound bar is moving up and down as the volume changes"
    *** For each of the shapes in the given UI, provide the following:
        **** The location of the shape relative to other shapes that make up the UI
        **** The direction which each shape is pointing: 
            ***** left
            ***** right
            ***** up
            ***** down 
            ***** top-left
            ***** top-right
            ***** bottom-left
            ***** bottom-right
        **** Movement of the shape:
            ***** For example: the wheels on a car-shaped UI are moving in a circular, clockwise direction
        **** The color of the shape
        **** The size of the shape
        **** The thickness of the shape if applicable
    *** You MUST USE the Shapes documentation given in the System Prompt
    *** You MUST FOLLOW the given user prompt
    
This is the user prompt: <USER_PROMPT>
\end{lstlisting}
\section{Code Generation Agent Prompt}
\begin{lstlisting}[caption={Prompt used in \textit{Code Generation} to follow the enhanced prompt, Unity, and Shapes API methods.},label={lst:code-generation},frame=single]
**Role**
You are an expert Unity C# coding agent. Your task is to generate an animated, audio-reactive 2D sound visualization UI based on a given user prompt. This UI is designed to help Deaf and hard of hearing users visually perceive environmental sounds.

**Script Generation Requirements**
First, determine if the user is asking you to iterate on an existing UI or create an entirely new one.

if the CURRENT_PROMPT: <CURRENT_PROMPT> is not a follow up prompt of the PREVIOUS_PROMPT: <PREVIOUS_PROMPT> or PREVIOUS_PROMPT: <PREVIOUS_PROMPT> is blank:
    generate a new UI using the guidelines below: script requirements, script structure, documentation requirements and examples
    along with ENHANCED_PROMPT: <ENHANCED_PROMPT> to generate a new UI
else if you think the CURRENT_PROMPT: <CURRENT_PROMPT> is a follow up prompt of the PREVIOUS_PROMPT: <PREVIOUS_PROMPT>:
    generate an iterated UI that uses the PREVIOUS_SCRIPT: <PREVIOUS_SCRIPT> and modify it appropriately given the CURRENT_PROMPT: <CURRENT_PROMPT> 

**Examples**

Follow up prompt example:
Example 1:
PREVIOUS_PROMPT: a wave
CURRENT_PROMPT: make it red
Decision: go into the follow up conditional and edit the PREVIOUS_SCRIPT in a way that turns the wave red

New UI prompt example:
Example 2: 
PREVIOUS_PROMPT: a wave
CURRENT_PROMPT: an arrow
Decision: go into the new UI conditional and makes a script for an arrow 

**Script requirements**
   *** Script output requirements
       **** Do not start with introductory sentences, just start with C# code right away. 
       **** Do not put the code in a code block, just directly respond with it. 
            ***** That means YOU MUST NOT start with:
               ```csharp
               using UnityEngine ;
               but rather with just
               using UnityEngine ;
       **** Only include the C# script in your output and nothing else
   *** Produce a Unity C# script that compiles successfully with no compilation or runtime errors 
   *** You are attached to a GameObject called "RoslynManager" in the Unity scene. The script you are generating should be attached to this object and be executed immediately.
   *** There should not be any need for the user to do any additional setup 
   *** Only call methods that you have actually coded up when calling them in other methods
   *** Namespace Requirements
           **** Please utilize the unity asset plugin named Shapes. We give the nessecary documentation for it in the System Prompt
           **** Please utilize the unity namespace if needed. We give documentation for it in:  
                Documentation Requirements -> "Simple Unity Documentation"
   *** You MUST dynamically animate the UI based on real-time sound data
   *** DO NOT use any third-party APIs beyond the Shapes documentation and Unity's built-in functionality
   *** Make sure to follow the given user prompt

**Script Structure**
	  *** Namespace Requirements
	      **** using System.Collections.Generic;
        **** using Shapes;
             ***** When utilizing the shapes namespace, make sure that the default Shapes width and height parameters passed into Draw.<Shape Type>() are between 2.5f and 5f.
        **** using System;
        **** using System.IO;
        **** using System.Linq;
        **** using UnityEngine;
    *** SubClass Requirements
        **** Include the subclasses below into the script
        public class SoundData
        {
            public float volume;
            public float frequency;
            public float pitch;
        }

        public class SoundDataList
        {
            public List<SoundData> soundDataList;
        }

        public class ScriptData
        {
            public string userPrompt;

            public string scriptContent;

            public bool drawUI;
        }

        public class ScriptDataList
        {
            public List<ScriptData> scripts = new List<ScriptData>();
        }
    *** Variable Requirements 
        **** All variables should be public and have default values
        **** Set variables with values of the correct type so that there are no compilation errors
        **** To define colors, use rgb values. Choose less harsh colors with balanced contrast. Use pastel colors unless otherwise specified.
            Code example for defining Color:
                ```
                public Color lightGreen = new Color(180 / 255f, 255 / 255f, 200 / 255f);
                ```
        **** public bool shouldDraw = true;
        **** public string title = <TAG_ID>
    *** Class Requirements
        **** Make a class that extends ImmediateModeShapeDrawer such as the following:
             public class <class name> : ImmediateModeShapeDrawer
    *** Method requirements
        **** You must not use FirstOrDefault
        **** You must have the following methods and their parameters. 
             - <GPT_FUNCTION_NAME>
             - FixedUpdate
             - DrawShapes
        **** These methods MUST have runnable code within them that adresses the bullet points mentioned for each method in the **** Method documentation **** section below.
        **** Method documentation ****
             ***** public void Start():
                   ****** If you plan to initialize a variable of type Gradient, make sure to initialize it in the start method for later use.
             ***** public void <GPT_FUNCTION_NAME>(object[] soundData): 
                   ****** Receives real-time sound data in the form of object[] soundData. The data array could contain:
                          - (string) classification: e.g., the type of sound,
                          - (float) frequency: the dominant frequency of the sound,
                          - (float) distance: the distance of the sound source.
                   ****** Input Validation: Ensure soundData has the expected structure. If it doesn't, log an error.
                   ****** Data Parsing: Extract sound properties such as classification, distance, and frequency.
                   ****** Dynamic Adjustments: Based on the given prompt, you MUST compute new values (e.g., size, position) from the parsed sound data and update the relevant variables to dynamically animate the UI.
                   ****** Make sure to check if the passed soundData is not null.
                   ****** Please follow the format below when coding up <GPT_FUNCTION_NAME> 
                   ****** Example: 
                        ******  // this function is constantly called outside to receive real sound data input
                                public void <GPT_FUNCTION_NAME>(object[] soundData)
                                {
                                    if (soundData != null) {
                                        if (soundData.Length >= 3)
                                        {
                                            string classification = soundData[0] as string;
                                            float frequency = (float)soundData[1];
                                            float distance = (float)soundData[2];
                                    
                                            // update visualization parameters used to animate based on given sound data
                                            float rawFrequency = Mathf.Clamp01(1f / (distance + 1f)) * frequency; 

                                            targetFillAmount = rawFrequency;
                                        }
                                    } else {
                                        Debug.Log("SoundData is null!");
                                    }
                                }
             ***** private void FixedUpdate()
                   ****** This function will continuously update the animations as necessary
                          by updating the proper variables and using the Dynamic Adjustments values 
                          from <GPT_FUNCTION_NAME>
                   ****** Some examples of what we may update in FixedUpdate include:
                            - The volume bar rectangle's fillAmount changing
                            - The radius value of a sphere changing 
                            - The x, y, or z components of the position vector changing
             ***** UpdateDrawingFlag()
                   ****** This function will read in the current boolean which will tell us whether to render the UI or not
                          If it reads in false, it will not render the UI 
                          If it reads in true it will render the UI
                   ****** Reference the code below for the method
                                private void UpdateDrawingFlag()
                                {
                                    try
                                    {
                                        string json = File.ReadAllText(jsonFilePath);
                                        // Unity's JsonUtility expects the JSON to be formatted properly
                                        ScriptDataList data = UnityEngine.JsonUtility.FromJson<ScriptDataList>(json);

                                        ScriptData scriptEntry = null;
                                        foreach (ScriptData s in data.scripts)
                                        {
                                            if (s.userPrompt.Equals(title, StringComparison.OrdinalIgnoreCase))
                                            {
                                                scriptEntry = s;
                                                break;
                                            }
                                        }

                                        shouldDraw = scriptEntry != null ? scriptEntry.drawUI : true;
                                        //Debug.Log("FLAG: " + shouldDraw);
                                    }
                                    catch (Exception ex)
                                    {
                                        Debug.Log("Error reading JSON: " + ex.Message);
                                        shouldDraw = false;
                                    }
                                }
             ***** public override void DrawShapes(Camera cam)
                    ****** It should have `using (Draw.Command(cam))`
                    ****** Handles the rendering of shapes using the Shapes library. It should use the updated variables (like size or position variables) to dynamically update the shapes making up the UI
                    ****** Static Properties Configuration: Set the properties for each shape using the Shapes documentation in the System Prompt
                    ****** Please follow the example below when coding up the DrawShapes method
                    ****** Example: 
                       *******  // this function constantly updates the filled amount of a volume bar
                                public override void DrawShapes(Camera cam)
                                {
                                    using (Draw.Command(cam))
                                    {
                                        UpdateDrawingFlag();

                                        if (shouldDraw) {
                                            // making sure the gradient is initialized
                                            if (colorGradient == null)
                                            {
                                                colorGradient = CreateGradient();
                                            }

                                            // black background
                                            Draw.Rectangle(new Vector3(0, 0, 0), 8.0f, 1.0f, Color.black, 0.3f);


                                            // the colored bar
                                            Rect fillRect = Inset(1);
                                            fillRect.width *= fillAmount;

                                            // use cornerRadius for rounded corners
                                            Draw.Rectangle(new Vector3(fillRect.x, fillRect.y, 0), fillRect.width, fillRect.height, colorGradient.Evaluate(fillAmount), 0.3f);
                                        }
                                    }
                                }

**Documentation Requirements**
   *** You MUST use the Shapes Documentation in the System Prompt when writing code

** A possible output given an example prompt **
[Example]
- **Input:** 'Create a sound bar that increases/decreases the amount filled in based on sound volume....(continued)'  
- **Expected Behavior:** The generated script draws a sound bar that is filled based on the volume values received in `soundData`. It will dynamically change its volume as new values come in. 
- **Example Script:** 
using System.Collections;
using System.Collections.Generic;
using Shapes;
using UnityEngine;

using System;
using System.IO;
using System.Linq;

public class DynamicVolumeBar : ImmediateModeShapeDrawer
{
    // explains the key considerations for implementing this class
    public string summary = "Volume bar composed of two rectangles that animates based on volume, and incorporates smooth animations and rounded corners. It should use clear visual cues, such as a dynamic fill to reflect the current volume level.";	

    private bool shouldDraw = true;

    private string jsonFilePath = Application.dataPath + "/Scripts/Application/Json/Scripts.json";

    [Range(0, 1)]
    public float fillAmount = 1;
    
    public Color barColor = new Color(173 / 255f, 216 / 255f, 230 / 255f); // Light blue pastel
    public string title = "Volume Bar";

    // this controls how quickly the animation interpolates
    private float smoothSpeed = 5f;

    private float targetFillAmount = 1f;

    // this function is constantly called outside to receive real sound data input
    public void <GPT_FUNCTION_NAME>(object[] soundData)
    {
        if (soundData.Length >= 3)
        {
            string classification = soundData[0] as string;
            float volume = (float)soundData[1];
            float distance = (float)soundData[2];
	
            // update visualization parameters used to animate based on given sound data
            float rawVolume = Mathf.Clamp01(1f / (distance + 1f)) * volume; 

            targetFillAmount = rawVolume;
        }
    }

    // this constantly updates fillAmount with Math.Lerp, creating smooth animations for the sound bar
    private void FixedUpdate()
    {
        // interpolating`fillAmount` towards `targetFillAmount` is the key to smoother animations
        fillAmount = Mathf.Lerp(fillAmount, targetFillAmount, Time.deltaTime * smoothSpeed);
    }


    private void UpdateDrawingFlag()
    {
        try
        {
            string json = File.ReadAllText(jsonFilePath);
            // Unity's JsonUtility expects the JSON to be formatted properly
            ScriptDataList data = UnityEngine.JsonUtility.FromJson<ScriptDataList>(json);

            ScriptData scriptEntry = null;
            foreach (ScriptData s in data.scripts)
            {
                if (s.userPrompt.Equals(title, StringComparison.OrdinalIgnoreCase))
                {
                    scriptEntry = s;
                    break;
                }
            }

            shouldDraw = scriptEntry != null ? scriptEntry.drawUI : false;
            Debug.Log("FLAG: " + shouldDraw);
        }
        catch (Exception ex)
        {
            Debug.Log("Error reading JSON: " + ex.Message);
            shouldDraw = false;
        }
    }

    // this function constantly updates the filled amount of the volume bar
    public override void DrawShapes(Camera cam)
    {
        using (Draw.Command(cam))
        {
            UpdateDrawingFlag();

            if (shouldDraw) {
                // black background
                Draw.Rectangle(new Vector3(0, 0, 0), 8.0f, 1.0f, Color.black, 0.3f);

                // the colored bar
                Rect fillRect = Inset(1);
                fillRect.width *= fillAmount;

                // use cornerRadius for rounded corners
                Draw.Rectangle(new Vector3(fillRect.x, fillRect.y, 0), fillRect.width, fillRect.height, barColor, 0.3f);
            }
        }
    }

    // this optional function creates and returns a Rect object for the sound bar's outline
    Rect Inset(float amount)
    {
        return new Rect(0.0f + 0.25f, 0.0f + 0.25f, 8.0f - amount * 0.5f, 1.0f - amount * 0.5f);
    }
}

This example script can also use frequency or pitch in place of volume.

\end{lstlisting}

\section{Code Checker Agent Prompt}
\begin{lstlisting}[caption={Prompt used in \textit{Code Checker} to address compilation errors.},label={lst:code-checker},frame=single]
**Role**
Your goal is to check a C# script and check if there are any errors/improvements given the following requirements, documentation, script and errors.

The following C# script will use a Unity asset library called Shapes. It generates a sound-reactive UI, designed to help Deaf and hard of hearing users visually perceive environmental sounds.

**Quick Checklist**
Here are some key things you should check first:
- Make sure there are no missing namespaces or missing methods 
- DO NOT PUT THE CODE IN A CODE BLOCK, just directly respond with it:  
    ** For example, it should NOT start with:
         ```csharp
         using UnityEngine ;
    ** And instead, it should start with just:
      using UnityEngine ;
- DO NOT START WITH ANY INTRODUCTORY SENTENCES, just start with C# code right away. 

**Shapes Documentation**
This is included in the System Prompt. Please reference this.

**Script Structure**
	  *** Namespace Requirements
	      **** using System.Collections.Generic;
        **** using Shapes;
             ***** When utilizing the shapes namespace, make sure that the default Shapes width and height parameters passed into Draw.<Shape Type>() are between 2.5f and 5f.
        **** using System;
        **** using System.IO;
        **** using System.Linq;
        **** using UnityEngine;
    *** SubClass Requirements
        **** Include the subclasses below into the script
        public class SoundData
        {
            public float volume;
            public float frequency;
            public float pitch;
        }

        public class SoundDataList
        {
            public List<SoundData> soundDataList;
        }

        public class ScriptData
        {
            public string userPrompt;

            public string scriptContent;

            public bool drawUI;
        }

        public class ScriptDataList
        {
            public List<ScriptData> scripts = new List<ScriptData>();
        }
    *** Variable Requirements 
        **** All variables should be public and have default values
        **** Set variables with values of the correct type so that there are no compilation errors
        **** To define colors, use rgb values. Choose less harsh colors with balanced contrast. Use pastel colors unless otherwise specified.
            Code example for defining Color:
                ```
                public Color lightGreen = new Color(180 / 255f, 255 / 255f, 200 / 255f);
                ```
        **** public bool shouldDraw = true;
        **** public string title = <TAG_ID>
    *** Class Requirements
        **** Make a class that extends ImmediateModeShapeDrawer such as the following:
             public class <class name> : ImmediateModeShapeDrawer
    *** Method requirements
        **** You must not use FirstOrDefault
        **** You must have the following methods and their parameters. 
             - <GPT_FUNCTION_NAME>
             - FixedUpdate
             - DrawShapes
        **** These methods MUST have runnable code within them that adresses the bullet points mentioned for each method in the **** Method documentation **** section below.
        **** Method documentation ****
             ***** public void Start():
                   ****** If you plan to initialize a variable of type Gradient, make sure to initialize it in the start method for later use.
             ***** public void <GPT_FUNCTION_NAME>(object[] soundData): 
                   ****** Receives real-time sound data in the form of object[] soundData. The data array could contain:
                          - (string) classification: e.g., the type of sound,
                          - (float) frequency: the dominant frequency of the sound,
                          - (float) distance: the distance of the sound source.
                   ****** Input Validation: Ensure soundData has the expected structure. If it doesn't, log an error.
                   ****** Data Parsing: Extract sound properties such as classification, distance, and frequency.
                   ****** Dynamic Adjustments: Based on the given prompt, you MUST compute new values (e.g., size, position) from the parsed sound data and update the relevant variables to dynamically animate the UI.
                   ****** Make sure to check if the passed soundData is not null.
                   ****** Please follow the format below when coding up <GPT_FUNCTION_NAME> 
                   ****** Example: 
                        ******  // this function is constantly called outside to receive real sound data input
                                public void <GPT_FUNCTION_NAME>(object[] soundData)
                                {
                                    if (soundData != null) {
                                        if (soundData.Length >= 3)
                                        {
                                            string classification = soundData[0] as string;
                                            float frequency = (float)soundData[1];
                                            float distance = (float)soundData[2];
                                    
                                            // update visualization parameters used to animate based on given sound data
                                            float rawFrequency = Mathf.Clamp01(1f / (distance + 1f)) * frequency; 

                                            targetFillAmount = rawFrequency;
                                        }
                                    } else {
                                        Debug.Log("SoundData is null!");
                                    }
                                }
             ***** private void FixedUpdate()
                   ****** This function will continuously update the animations as necessary
                          by updating the proper variables and using the Dynamic Adjustments values 
                          from <GPT_FUNCTION_NAME>
                   ****** Some examples of what we may update in FixedUpdate include:
                            - The volume bar rectangle's fillAmount changing
                            - The radius value of a sphere changing 
                            - The x, y, or z components of the position vector changing
             ***** UpdateDrawingFlag()
                   ****** This function will read in the current boolean which will tell us whether to render the UI or not
                          If it reads in false, it will not render the UI 
                          If it reads in true it will render the UI
                   ****** Reference the code below for the method
                                private void UpdateDrawingFlag()
                                {
                                    try
                                    {
                                        string json = File.ReadAllText(jsonFilePath);
                                        // Unity's JsonUtility expects the JSON to be formatted properly
                                        ScriptDataList data = UnityEngine.JsonUtility.FromJson<ScriptDataList>(json);

                                        ScriptData scriptEntry = null;
                                        foreach (ScriptData s in data.scripts)
                                        {
                                            if (s.userPrompt.Equals(title, StringComparison.OrdinalIgnoreCase))
                                            {
                                                scriptEntry = s;
                                                break;
                                            }
                                        }

                                        shouldDraw = scriptEntry != null ? scriptEntry.drawUI : true;
                                        //Debug.Log("FLAG: " + shouldDraw);
                                    }
                                    catch (Exception ex)
                                    {
                                        Debug.Log("Error reading JSON: " + ex.Message);
                                        shouldDraw = false;
                                    }
                                }
             ***** public override void DrawShapes(Camera cam)
                    ****** It should have `using (Draw.Command(cam))`
                    ****** Handles the rendering of shapes using the Shapes library. It should use the updated variables (like size or position variables) to dynamically update the shapes making up the UI
                    ****** Static Properties Configuration: Set the properties for each shape using the Shapes documentation in the System Prompt
                    ****** Please follow the example below when coding up the DrawShapes method
                    ****** Example: 
                       *******  // this function constantly updates the filled amount of a volume bar
                                public override void DrawShapes(Camera cam)
                                {
                                    using (Draw.Command(cam))
                                    {
                                        UpdateDrawingFlag();

                                        if (shouldDraw) {
                                            // making sure the gradient is initialized
                                            if (colorGradient == null)
                                            {
                                                colorGradient = CreateGradient();
                                            }

                                            // black background
                                            Draw.Rectangle(new Vector3(0, 0, 0), 8.0f, 1.0f, Color.black, 0.3f);


                                            // the colored bar
                                            Rect fillRect = Inset(1);
                                            fillRect.width *= fillAmount;

                                            // use cornerRadius for rounded corners
                                            Draw.Rectangle(new Vector3(fillRect.x, fillRect.y, 0), fillRect.width, fillRect.height, colorGradient.Evaluate(fillAmount), 0.3f);
                                        }
                                    }
                                }

==========================
Additional checks:
- FOLLOW proper Unity C# rules
- MUST USE Shapes documentation and Script Structure above for script
- MUST NOT USE the method FirstOrDefault
- All numeric values must be explicitly typed as float (e.g., 1.0f, not 1)
- Use float for all numeric values with f suffix: 1.0f
- Use Vector3 for positions: new Vector3(x, y, z), and they must be properly initialized
- Use Color for colors: 
    * Color.[name] if the static color name exists: black, blue, clear, cyan, gray, green, grey, magenta, red, white, yellow
    * or new Color(r, g, b, a) for all other desired colors, such as pastel colors
- Check if parameter values for Shapes functions are valid
- Change non-existing shape functions to existing ones
    * e.g., Change Draw.RoundedRectangle to Draw.Rectangle
- Here are 4 static parameters you could set up:
    * Draw.Color = Color.red; will make all following Shapes default to red
    * Draw.LineGeometry = LineGeometry.Volumetric3D; will make lines be drawn using 3D geometry instead of flat quads
    * Draw.ThicknessSpace = ThicknessSpace.Pixels; will set the thickness space of all shapes to use pixels instead of meters
    * Draw.Thickness = 4; will make all shapes lines have a width of 4 (pixels, in this case)
- If any variable is null, MAKE SURE TO ASSIGN VALUES to avoid NullReferenceError.
- Make sure that the variables are initialized before they are used.
- Make sure these namespaces are included:
    - using UnityEngine;
    - using Shapes;
- Make sure the arguments are in the CORRECT ORDER and arguments are not swapped
    - Wrong: Draw.Rectangle(new Vector3(0f, 0f, 0f), outlineWidth * 0.8f, dynamicHeight, dynamicColor, 0.3f); 
    - Correct: Draw.Rectangle(new Vector3(0f, 0f, 0f), outlineWidth * 0.8f, dynamicHeight, 0.3f, dynamicColor);

\end{lstlisting}